\documentclass{ws-acs}

\begin{document}

\markboth{R. V. Sol{\'e}, R. Pastor-Satorras, E. Smith, and T. Kepler}
{A model of Large-scale Proteome Evolution}

\catchline

\title{A MODEL OF LARGE-SCALE PROTEOME EVOLUTION}

\author{Ricard V. Sol{\'e}$^{1,2,3}$, Romualdo Pastor-Satorras$^1$,
  Eric Smith$^2$, and Thomas B. Kepler$^2$} 

\address{$^1$ICREA-Complex Systems Research Group, FEN\\
    Universitat Polit{\`e}cnica de Catalunya, Campus Nord B4,
    08034 Barcelona, Spain}
\address{$^2$Santa Fe Institute, 1399 Hyde Park Road, New Mexico
  87501, USA}
\address{$^3$NASA-associated Astrobiology Institute, INTA/CSIC,
  Carr. del Ajalvir Km4, Madrid, Spain}

\maketitle

\begin{abstract}
  The next step in the understanding of the genome organization, after
  the determination of complete sequences, involves proteomics. The
  proteome includes the whole set of protein-protein interactions, and
  two recent independent studies have shown that its topology displays
  a number of surprising features shared by other complex networks,
  both natural and artificial. In order to understand the origins of
  this topology and its evolutionary implications, we present a simple
  model of proteome evolution that is able to reproduce many of the
  observed statistical regularities reported from the analysis of the
  yeast proteome. Our results suggest that the observed patterns can
  be explained by a process of gene duplication and diversification
  that would evolve proteome networks under a selection pressure,
  favoring robustness against failure of its individual components.
  
\end{abstract}

\section{Introduction}

The genome is one of the most fascinating examples of the importance
of emergence from network interactions. The recent sequencing of the
human genome \cite{lander01,Venter01} revealed some
unexpected features and confirmed that {\em ``the sequence is only the
  first level of understanding of the genome''} \cite{Venter01}.
The next fundamental step beyond the determination of the genome
sequence involves the study of the properties of the proteins the
genes encode, as well as their interactions \cite{fields01}.  Protein
interactions play a key role at many different levels and its failure
can lead to cell malfunction or even apoptosis, in some cases
triggering neoplastic transformation. This is the case, for example,
of the feedback loop between two well-known proteins, MDM2 and p53: in
some types of cancers, amplification of the first (an oncoprotein)
leads to the inactivation of p53, a tumor-suppressor gene that is
central in the control of the cell cycle and death \cite{Wu93}.

Understanding the specific details of protein-protein interactions is
an essential part of our understanding of the proteome, but a
complementary approach is provided by the observation that
network-like effects play also a key role. Using again p53 as an
example, this gene is actually involved in a large number of
interaction pathways dealing with cell signaling, the maintenance of
genetic stability, or the induction of cellular differentiation
\cite{Vogelstein00}. The failure in p53, as when a highly
connected node in the Internet breaks \cite{barabasi00}, has severe
consequences.

Additional insight is provided by the observation
that in many cases the total suppression of a given gene in a given
organism leads to a small phenotypic effect or even no effect at all 
\cite{ross99,wagner00}. These observations support 
the idea that, although some genes might
play a key role and their suppression is lethal, many others can be
replaced in their function by some redundancy implicit in the network
of interacting proteins. 

Protein-protein interaction maps have been studied, at different
levels, in a variety of organisms including viruses
\cite{bartel96,flajolet00,McCraith00}, prokaryotes \cite{rain01},
yeast \cite{ito00}, and multicellular organisms such as {\em C.
  elegans} \cite{walhout00}. Most previous studies have used the
so-called two-hybrid assay \cite{fromont97} based on the properties of
site-specific transcriptional activators.  Although differences exist
between different two-hybrid projects \cite{hazbun01} the statistical
patterns used in our study are robust.

\begin{figure}[t]

\centerline{\psfig{file=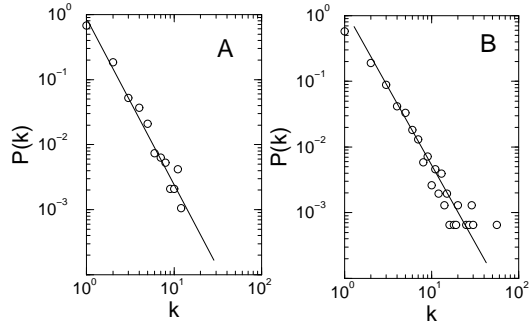,width=7cm}}

\vspace*{8pt}
\caption{Degree distributions for two different data sets 
  from the Yeast proteome: A: Ref.~\protect\cite{wagner01}; B:
  Ref.~\protect\cite{jeong01}.  Both distributions display scaling
  behavior in their degree distribution $P(k)$, i.e. $P(k) \sim
  k^{-\gamma}$, a sharp cut-off for large $k$ and very small average
  connectivities: $\bar{K}_A=1.83$ (total graph) and $\bar{K}_B=2.3$
  (giant component), respectively. The slopes are $\gamma_A \approx 2.5 \pm 0.15$
  and $\gamma_B \approx 2.4 \pm 0.21$.}
\label{fig:proteomefig1}
\end{figure}

Recent studies have revealed a surprising result: the protein-protein
interaction networks in the yeast {\em Saccharomyces cerevisiae} share
some universal features with other complex networks \cite{strog01}.
These studies actually offer the first global view of the proteome
map. These are very heterogeneous networks: The probability $P(k)$
that a given protein interacts with other $k$ proteins is given by a
power law, i.e. $P(k) \sim k^{-\gamma}$ with $\gamma \approx 2.5$ (see figure
\ref{fig:proteomefig1}), with a sharp cut-off for large $k$.
This distribution is thus very different from the Poissonian shape
expected from a simple (Erdos-Renyi) random graph \cite{bollobas,kauffman93}.
Additionally, these maps also display the so-called small-world (SW) effect:
they are highly clustered (i.e.  each node has a well-defined
neighborhood of ``close'' nodes) but the minimum distance between any
two randomly chosen nodes in the graph is short, a characteristic
feature of random graphs \cite{watts98}.

As shown in previous studies \cite{barabasi00} this type of networks
is extremely robust against random node removal but also very fragile
when removal is performed selectively on the most connected nodes.  SW
networks appear to be present in a wide range of systems, including
artificial ones \cite{barab99,amaral,ferrer01,pvv} and also in neural
networks \cite{watts98,stephan}, metabolic pathways \cite{fell00,jeong00,wagnerfell} (see
also \cite{Onzonnis}),
even in human language organization \cite{ferrer01b}. The implications
of these topologies are enormous also for our understanding of
epidemics \cite{pv01a,lloyd01}.

The experimental observations on the proteome map can be summarized as
follows:
\begin{itemize}
\item[(1)] The proteome map is a sparse graph, with a small average number
  of links per protein. In \cite{wagner01} an average connectivity
$\bar{K} \sim 1.9-2.3$ was reported for the proteome map of {\em
S. cerevisiae}. 
 This observation is also consistent with the
  study of the global organization of the {\em E. coli} gene network
  from available information on transcriptional regulation
  \cite{Thieffry98}.

\item[(2)] It exhibits a SW pattern, different from the properties
  displayed by purely random (Poissonian) graphs.

\item[(3)] The degree distribution of links follows a power-law with a
  well-defined cut-off. To be more precise, Jeong {\em et al.}
  \cite{jeong01} reported a functional form for the degree
  distribution of {\em S. cerevisiae}
  \begin{equation}
    P(k) \simeq (k_0 + k)^{-\gamma} e^{-k/k_c}.
    \label{eq:2}
  \end{equation}
  A best fit of the real data to this form yields a degree exponent
  $\gamma \approx 2.5$ and a cut-off $k_c\approx20$. This could have adaptive
  significance as a source of robustness against mutations.

\end{itemize}

In this paper we present a model of proteome evolution aimed at
capturing the main properties exhibited by protein networks. The basic
ingredients of the model are gene duplication plus re-wiring of the
protein interactions, two elements known to be the essential driving
forces in genome evolution \cite{ohono70}.  The model does not include
functionality or dynamics of the proteins involved, but it is a
topologically-based approximation to the overall features of the
proteome graph and intends to capture some of the generic
features of proteome evolution.

During the completion of this work we became aware of a paper by
V{\'a}zquez et al., Ref.~\cite{vazquez}, in which a related model of
proteome evolution, showing multifractal connectivity properties, is
described and analyzed.

\section{Proteome growth model}

Here we restrict our rules to single-gene duplications, which occur in
most cases due to unequal crossover \cite{ohono70}, plus re-wiring. Multiple
duplications should be considered in future extensions of these
models: molecular evidence shows that even whole-genome duplications
have actually occurred in {\em S. cerevisiae} \cite{Wolfe97} (see also
Ref.~\cite{wagner94}). Re-wiring has also been used in dynamical
models of the evolution of robustness in complex organisms
\cite{Bornholdt00}.

It is worth mentioning that the study of metabolic networks provides
some support to the rule of {\em preferential attachment} \cite{barab99} as a candidate
mechanism to explain the origins of the scale-free
topology. Scale-free graphs are easily obtained in a growing network
provided that the links to new nodes are made preferentially from
nodes that already have many links. A direct consequence is that
vertices with many connections are those that have been incorporated
early. This seems to be plausible in the early history of metabolic
nets, and this view is supported by some available evidence
\cite{wagnerfell}. A similar argument can be made with proteome maps,
since there are strong connections between the evolution of metabolic
pathways and genome evolution, and other scenarios have also been
proposed, including optimization \cite{ferrersole}. Here we do not
consider preferential attachment rules, although future studies should
explore the possible contributions of different mechanisms to the
evolution of network biocomplexity. In this context, new integrated
analyses of cellular pathways using microarrays and quantitative
proteomics \cite{ideker} will help to obtain a more detailed picture of how these
networks are organized.

\begin{figure}[t]

\centerline{\psfig{file=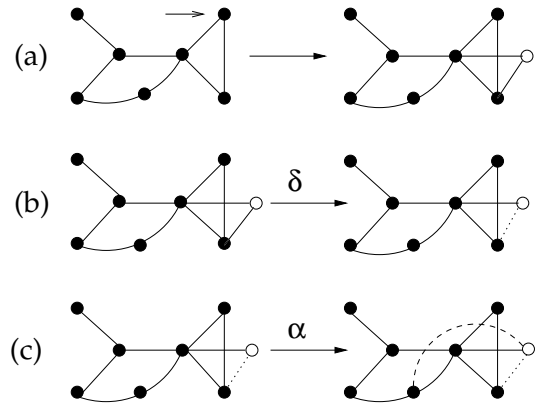,width=7cm}}

\vspace{8pt}
\caption{Growing network by duplication of nodes. First (a) duplication 
occurs after randomly selecting a node (arrow). The links from the 
newly created node (white) now can experience deletion (b) and new links  
can be created (c); these events occur with probabilities $\delta$ and 
$\alpha$, respectively.} 
\label{fig:proteorulesPRS}
\end{figure}

The proteome graph at any given step $t$ (i.e. after $t$ duplications)
will be indicated as $\Omega_p (t)$. The rules of the model, summarized in
figure \ref{fig:proteorulesPRS}, are implemented as follows.  Each
time step: (a) one node in the graph is randomly chosen and
duplicated; (b) the links emerging from the new generated node are
removed with probability $\delta$; (c) finally, new links (not previously
present) can be created between the new node and all the rest of the
nodes with probability $\alpha$. Step (a) implements gene duplication, in
which both the original and the replicated proteins retain the same
structural properties and, consequently, the same set of interactions.
The rewiring steps (b) and (c) implement the possible mutations of the
replicated gene, which translate into the deletion and addition of
interactions, with different probabilities.

Since we have two free parameters, we should first constrain their
possible values by using the available empirical data.  As a first
step, we can estimate the asymptotic average connectivity exhibited by
the model in a mean-field approximation (see also
Ref.~\cite{vazquez}). Let us indicate by $\bar{K}_N$ the average
connectivity of the system when it is composed by $N$ nodes.  It is
not difficult to see that the increase in the average connectivity
after one iteration step of the model is proportional to
\begin{equation}
  \frac{d \bar{K}_N}{d N} \simeq \bar{K}_{N+1} -\bar{K}_{N} = \frac{1}{N}
  \left[ \bar{K}_N - 2 \delta \bar{K}_N +  2 \alpha (N -\bar{K}_N) \right].
  \label{eq:3}
\end{equation}
The first term accounts for the duplication of one node, the second
represents the average elimination of $\delta \bar{K}_N$ links emanating
from the new node, and the last term represents the addition of $\alpha (N
-\bar{K}_N)$ new connections pointing to the new
node. Eq.~(\ref{eq:3}) is a linear equation which easily solved,
yielding
\begin{equation}
  \bar{K}_N=\frac{\alpha N}{\alpha+\delta} + \left( \bar{K}_1 - \frac{\alpha}{\alpha+\delta}
  \right) N^\Gamma, 
\end{equation}
where $\Gamma=1-2 \alpha -2 \delta$ and $\bar{K}_1$ is the initial average
connectivity of the system. This solution leads to an increasing
connectivity through time.  In order to have a finite $\bar{K}$ in the
limit of large $N$, we must impose the condition $\alpha = \beta/N$, where
$\beta$ is a constant. That is, the rate of addition of new links (the
establishment of new viable interactions between proteins) is
inversely proportional to the network size, and thus much smaller than
the deletion rate $\delta$, in agreement with the rates observed in
\cite{wagner01}. In this case, for large $N$, we get
\begin{equation}
  \frac{d \bar{K}_N}{d N} = \frac{1}{N} (1 - 2 \delta) \bar{K}_N +
  \frac{2\beta}{N}. 
\end{equation}
The solution of this equation is
\begin{equation}
  \bar{K}_N=\frac{2 \beta}{2\delta -1} + \left( \bar{K}_1 - \frac{2 \beta}{2\delta-1}
  \right) N^{1-2\delta}.
\end{equation}
For $\delta>1/2$ a finite connectivity is reached,
\begin{equation}
\bar{K} \equiv \bar{K}_\infty = \frac{2 \beta}{2\delta -1}.
\label{eq:boundary}
\end{equation}
The previous expression imposes the boundary condition $\delta>1/2$,
necessary in order to obtain a well-defined limiting average
connectivity. Eq.~(\ref{eq:boundary}), together with the experimental
estimates of $\bar{K} \sim 1.9 -2.3$, allows to set a first restriction
to the parameters $\beta$ and $\delta$. Imposing $\bar{K} = 2$, we are led to the relation
\begin{equation}
\beta = 2 \delta -1.
\label{eq:lligam1}
\end{equation}
Moreover, estimations of addition and deletion rates $\alpha$ and $\delta$
from yeast \cite{wagner01} give a ratio $\alpha/\delta \leq 10^{-3}$.  For
proteomes of size $N \sim 10^3$, as in the case of the yeast, this leads
to $\beta/\delta \leq 10^{-3} N \sim 1$. Using the safe approximation $\beta/ \delta \approx
0.25$, together with the constraint~(\ref{eq:boundary}), we obtain the
approximate values
\begin{equation}
  \delta =  0.58, \qquad \beta =  0.16.
  \label{eq:1}
\end{equation}
which will be used through the rest of the paper. 

Simulations of the model start form a connected ring of $N_0=5$ nodes,
and proceed by iterating the rules until the desired network size is
achieved.

\section{Results}

\begin{figure}[t]

\centerline{{\large A)}\psfig{file=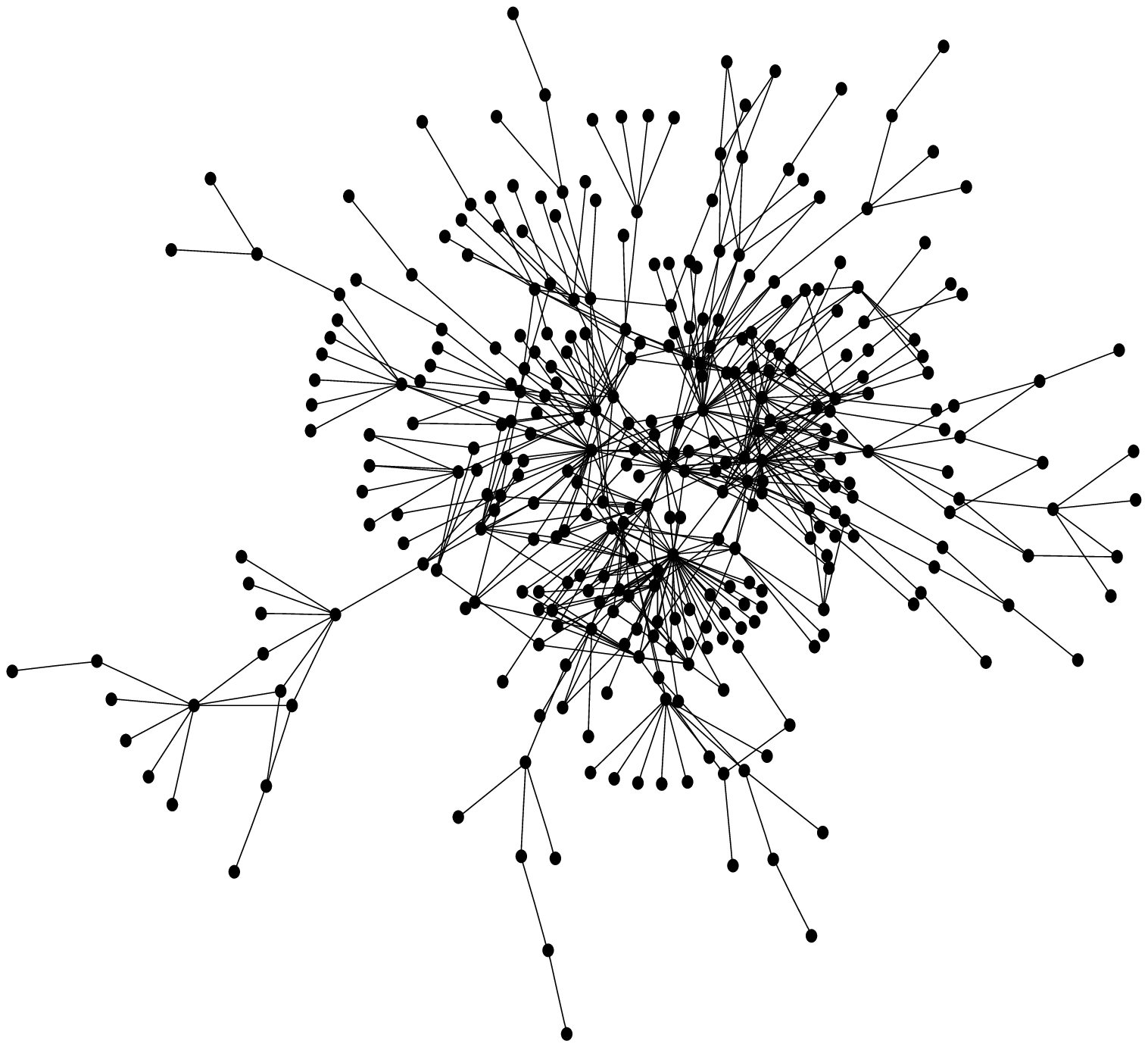,width=5cm}\hspace*{1cm}
{\large B)}\psfig{file=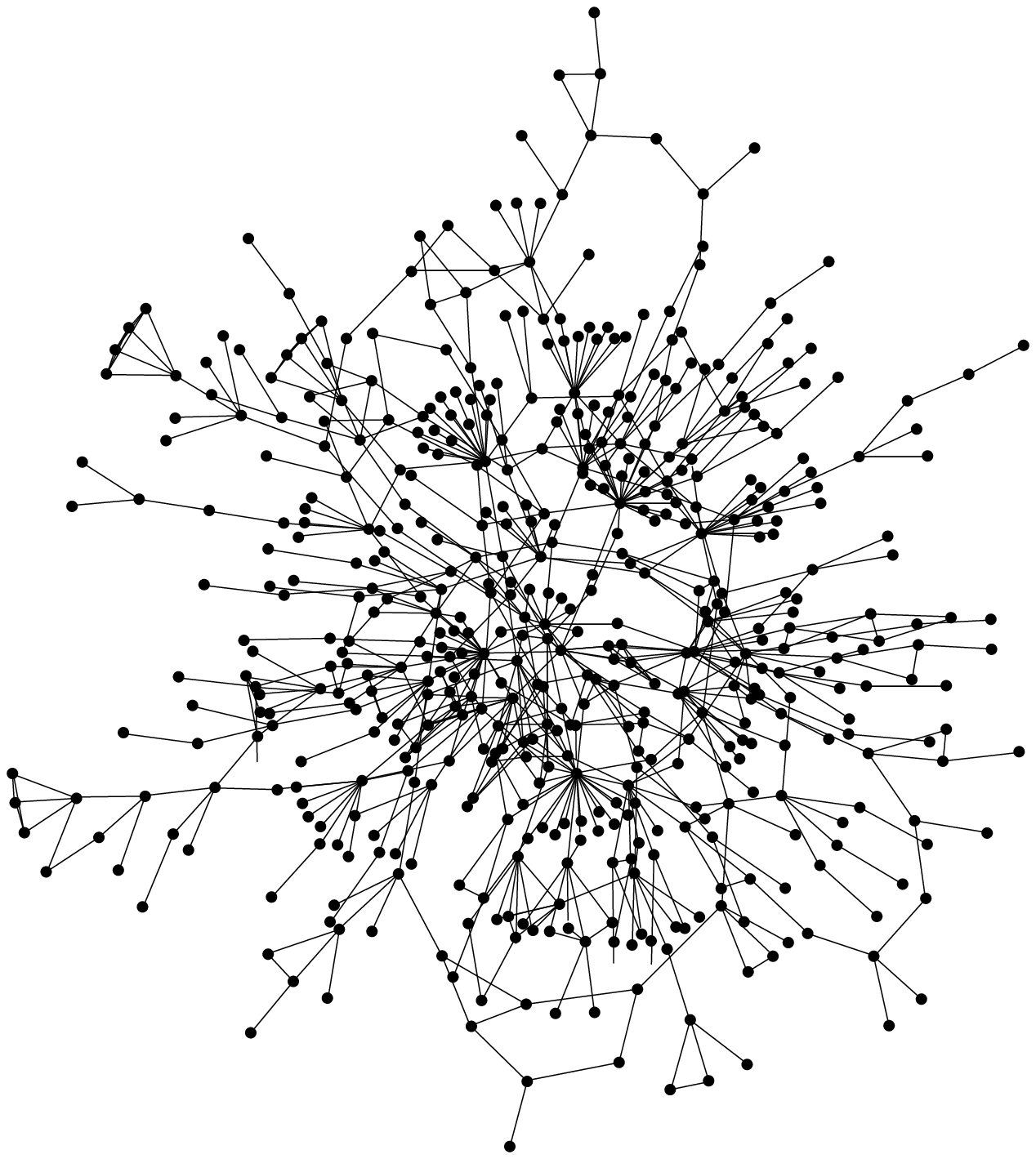,width=5cm}}

\vspace{8pt}
\caption{A) An example of a small proteome interaction map (giant component, 
  $\Omega_{\infty}$) generated by the model with $N=10^3$, $\delta=0.58$, and
  $\beta=0.16$. B) Real yeast proteome map obtained from the MIPS
  database \protect\cite{mewes99}.
   We can observe the close similitude between the
  real map and the output of the model.}
\label{fig:proteomegraf}
\end{figure}

\begin{figure}[t]

\centerline{\psfig{file=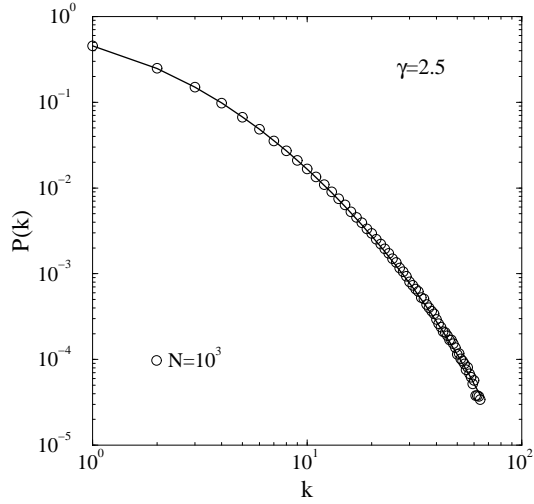,width=7cm}}
\vspace{8pt}
\caption{Degree distribution $P(k)$ for the model, averaged over
  $10^4$ networks of size $N=10^3$. The distribution shows a
  characteristic power law behavior, with exponent $\gamma = 2.5 \pm 0.1$
  and an exponential cut-off $k_c\simeq28$.}
\label{fig:proteomedist}
\end{figure}

Computer simulations of the proposed model reproduce many of the
regularities observed in the real proteome data. As an example of the
output of the model, in figure \ref{fig:proteomegraf}A we show an
example of the giant component $\Omega_{\infty}$ (the largest cluster
of connected proteins) of a realization of the model
with $N=10^3$ nodes. This figure clearly resembles the giant component
of real yeast networks, as we can see comparing with figure
\ref{fig:proteomegraf}B\footnote{Figure kindly provided by W. Basalaj
  (see http://www.cl.cam.uk/~wb204/GD99/\#Mewes).}, and we can
appreciate the presence of a few highly connected hubs plus many nodes
with a relatively small number of connections. The size of the giant
component for $N=10^3$, averaged of $10^4$ networks, is $\vert \Omega_{\infty}
\vert = 472 \pm 87$, in good agreement with Wagner's data $\vert
\Omega^W_{\infty} \vert = 466$ for a yeast with a similar total number of
proteins (the high variance in our result is due to the large
fluctuations in the model for such small network size $N$). On the
other hand, in figure \ref{fig:proteomedist} we plot the connectivity
$P(k)$ obtained for networks of size $N=10^3$. In this figure we
observe that the resulting connectivity distribution can be fitted to
a power-law win an exponential cut-off, of the form given by
Eq.~(\ref{eq:2}), with parameters $\gamma=2.5\pm0.1$ and $k_c\simeq28$, in good
agreement with the measurements reported in Refs~\cite{wagner01}
and~\cite{jeong01}.

An additional observation from Wagner's study of the yeast proteome is
the presence of SW properties. We have found also similar topological
features in our model, using the considered set of parameters.  The
proteome graph is defined by a pair $\Omega_p=(W_p, E_p)$, where $W_p=\{ p_i
\}, (i=1, ..., N)$ is the set of $N$ proteins and $E_p=\{ \{p_i, p_j\} \}$
is the set of edges/connections between proteins.  The {\em adjacency
  matrix} $\xi_{ij}$ indicates that an interaction exists between
proteins $p_i, p_j \in \Omega_p$ ($\xi_{ij} = 1$) or that the interaction is
absent ($\xi_{ij} = 0$).  Two connected proteins are thus called {\em
  adjacent} and the {\em degree} of a given protein is the number of
edges that connect it with other proteins.

The SW pattern can be detected from the analysis of two basic
statistical quantities: the {\em clustering coefficient} $C_v$ and the
{\em average path length} $L$. Let us consider the adjacency matrix and
indicate by $\Gamma_i=\{ p_i \,\vert \, \xi_{ij} = 1 \}$ the set of nearest
neighbors of a protein $p_i \in W_p$. The clustering coefficient for
this protein is defined as the number of connections between the
proteins $p_j \in \Gamma_i$ \cite{watts98}. Denoting
\begin{equation}
{\cal L}_i = \sum_{j=1}^{N} \xi_{ij} \left [ \sum_{k \in \Gamma_i} \xi_{jk} \right ],
\end{equation}
we define the clustering coefficient of the $i$-th protein as
\begin{equation}
  c_v(i)= \frac{2 {\cal L}_i}{k_i (k_i-1)},
\end{equation}
where $k_i$ is the connectivity of the $i$-th protein. The clustering
coefficient is defined as the average of $c_v(i)$ over all the
proteins,
\begin{equation}
C_v= {1 \over N} \sum_{i=1}^{N} c_v(i),
\end{equation}
and it provides a measure of the average fraction of pairs of
neighbors of a node that are also neighbors of each other.

\begin{table}[t]
\tbl{Comparison between the observed regularities in the yeast proteome
\protect\cite{wagner01}, the model predictions with $N=10^3$, $\delta=0.58$ and
$\beta=0.16$, and a random network with the same size and average
connectivity as the model. The quantities $X$ represent averages over
the whole graph; $X^g$ represent averages over the giant component.}
{\begin{tabular}{@{}cccc@{}}\toprule
    & Yeast proteome &  Network model & Random network\\ \colrule
    $\bar{K}$ & $1.83$  & $2.2 \pm 0.5$   & $2.00\pm0.06$ \\ 
    $\bar{K}^g$ & $2.3$  & $4.3 \pm 0.5$  & $2.41 \pm 0.05$ \\
    $\gamma$ & $2.5$  & $2.5 \pm 0.1$  & --- \\ 
    %$\gamma^g$ & $2.5$  & $2.5 \pm 0.2$  & --- \\ \hline
    $\vert\Omega_{\infty} \vert$  &  $466$  &  $472 \pm 87$ &  $795 \pm 22$ \\ 
    $C_v^g$  &  $2.2 \times 10^{-2}$  &  $1.0 \times 10^{-2}$  & $1.5 \times 10^{-3}$ \\ 
    $L^g$    &   $7.14$  &  $5.1\pm 0.5$ & $9.0 \pm 0.4$ \\ \botrule
  \end{tabular}}
\label{table1}
\end{table}

The average path length $L$ is defined as follows: Given two proteins $p_i,
p_j \in W_p$, let $L_{min}(i,j)$ be the minimum path length connecting
these two proteins in $\Omega_p$.  The average path length $L$ will be:
\begin{equation}
L = {2 \over N(N-1) } \sum_{i<j}^{N} L_{min}(i,j)
\end{equation}

Random graphs, where nodes are randomly connected with a given
probability $p$ \cite{bollobas}, have a clustering coefficient
inversely proportional to the network size, $C_v^{\rm rand} \approx
\bar{K}/N$, and an average path length proportional to the logarithm
of the network size, $L^{\rm rand} \approx \log N / \log \bar{K}$.  At the
other extreme, regular lattices with only nearest-neighbor connections
among units are typically clustered and exhibit long average paths.
Graphs with SW structure are characterized by a high clustering with
$C_v \gg C_v^{rand}$, while possessing an  average path comparable with a
random graph with the same connectivity and number of nodes.

In Table~\ref{table1} we report the values of $\bar{K}$, $\gamma$,
$\vert\Omega_{\infty} \vert$, $C_v$, and $L$ for our model, compared with the
values reported for the yeast {\em S. cerevisiae}
\cite{jeong01,wagner01}, and the values corresponding to a random
graph with size and connectivity comparable with both the model and
the real data. Except the average connectivity of the giant component,
which is slightly larger for the model, all the magnitudes for the
model compare quite well with the values measured for the yeast. On
the other hand, the values obtained for a random graph support the
conjecture of the SW properties of the protein network put forward in
Ref.~\cite{wagner01}.

\section{Discussion}

The analysis of complex biological networks in terms of random graphs
is not new. Early work suggested that the understanding of some
general principles of genome organization might be the result of
emergent properties within random networks of interacting units
\cite{kauffman62,kauffman93}. An important difference emerges, however, from
the new results about highly heterogeneous networks: the topological
organization of metabolic and protein graphs is very different from
the one expected under totally random wiring and as a result of their
heterogeneity, new qualitative phenomena emerge (such as the
robustness against mutation). This supports the view that cellular
functions are carried out by networks made up by many species of
interacting molecules and that networks of interactions might be at
least as important as the units themselves \cite{hartwell99,salazar00}.

Our study has shown that the macroscopic features exhibited by the proteome 
are also present in our simple model. This is surprising, since it is 
obvious that different proteins and protein interactions play different 
roles and operate under very different time scales and our model 
lacks such specific properties, dynamics or explicit functionality. Using 
estimated rates of addition and deletion of protein interactions 
as well as the average connectivity of the yeast proteome, we accurately 
reproduce the available statistical regularities exhibited by the real 
proteome. In this context, although data from yeast might involve
several sources of bias, it has been shown that the same type of
distribution is observable in other organisms, such as the protein
interaction map of the human gastric pathogen {\em Helicobacter
pylori} or in the p53
network (Jeong and Barab\'asi, personal communication). 

These results suggest that the global organization of protein
interaction maps can be explained by means of a simple process of gene
duplication plus diversification.  These are indeed the mechanisms
known to be operating in genome evolution (although the magnitude of
the duplication event can be different). One important point to be
explored by further extensions of this model is the origin of the
specific parameters used. The use of evolutionary algorithms and
optimization procedures might provide a consistent explanation of the
particular values observed and their relevance in terms of
functionality. A different source of validation of our model might be
the study of proteome maps resulting from the evolution of resident
genomes \cite{andersson}: the genomes of endosymbionts and cellular
organelles display an evolutionary degradation that somehow describe
an inverse rule of proteome reduction. Reductive evolution can be
almost extreme, and available data of resident proteomes might help to
understand how proteome maps get simplified under the environmental
conditions defined by the host genome. If highly connected nodes play
a relevant role here, perhaps resident genomes shrink by loosing
weakly connected nodes first.

Most of the classic
literature within this area deal with the phylogenetic consequences of
duplication and do not consider the underlying dynamics of
interactions between genes. We can see, however, that the final
topology has nontrivial consequences: this type of scale-free network
will display an extraordinary robustness against random removal of
nodes \cite{barabasi00} and thus it can have a selective role.
But an open question arises: is the scale-free organization observed
in real proteomes a byproduct of the pattern of duplication plus
rewiring (perhaps under a low-cost constraint in wiring) and thus we
have ``robustness for free''? The alternative is of course a
fine-tuning of the process in which selection for robustness has been
obtained by accepting or rejecting single changes. Further model
approximations and molecular data might provide answers to these
fundamental questions.

\section*{Acknowledgements}
The authors thank J. Mittenthal, R. Ferrer, J. Montoya, S. Kauffman
and A. Wuensche for useful discussions. This work has been supported
by a grant PB97-0693 and by the Santa Fe Institute (RVS). RPS
acknowledges financial support from the Ministerio de Ciencia y
Tecnolog{\'\i}a (Spain).

\end{document}